\documentclass[aps,prl,twocolumn,groupedaddress]{revtex4}
\usepackage[squaren]{SIunits}
\usepackage{amsmath}
\usepackage{epsfig}

\begin{document}

\def\d{{\rm d}}
\def\eps{\varepsilon}
\def\lp{\left. }
\def\rp{\right. }
\def\lr{\left( }
\def\rr{\right) }
\def\le{\left[ }
\def\re{\right] }
\def\lg{\left\{ }
\def\rg{\right\} }
\def\lb{\left| }
\def\rb{\right| }
\def\beq{\begin{equation}}
\def\eeq{\end{equation}}
\def\bea{\begin{eqnarray}}
\def\eea{\end{eqnarray}}
\def\mr{\,\mathrm{}}

\preprint{DESY 15-157}
\preprint{MS-TP-15-13}
\title{Next-to-next-to-leading order contributions to inclusive jet production
 in deep-inelastic scattering and determination of $\alpha_s$}
\author{Thomas Biek\"otter$^a$}
\author{Michael Klasen$^a$}
\email[]{michael.klasen@uni-muenster.de}
\author{Gustav Kramer$^b$}
\affiliation{$^a$ Institut f\"ur Theoretische Physik, Westf\"alische
 Wilhelms-Universit\"at M\"unster, Wilhelm-Klemm-Stra\ss{}e 9, D-48149 M\"unster,
 Germany\\
 $^b$ II.\ Institut f\"ur Theoretische Physik, Universit\"at Hamburg, Luruper
 Chaussee 149, D-22761 Hamburg, Germany}
\date{\today}
\begin{abstract}
We present the first calculation of inclusive jet production in deep-inelastic
scattering with approximate next-to-next-to-leading order (aNNLO) contributions,
obtained from a unified threshold resummation formalism. The leading coefficients
are computed analytically. We show that the aNNLO contributions
reduce the theoretical prediction for jet production in deep-inelastic scattering,
improve the description of the final HERA data in particular at high photon
virtuality $Q^2$ and increase the central fit value of the strong coupling constant.
\end{abstract}
\maketitle

\vspace*{-84mm}
\noindent DESY 15-157 \\
\noindent MS-TP-15-13 \\
\vspace*{61mm}


\section{Introduction}

The HERA collider, which operated at DESY from 1992 to 2007, has
produced many important physics results, first of all perhaps the most
precise determinations to date of the quark and gluon densities in the
proton from single experiments (H1, ZEUS) \cite{Adloff:2000qk,%
Chekanov:2001qu} and their combined data sets \cite{Aaron:2009aa}. These
data, taken in deep-inelastic electron-proton scattering, are
complemented by a wealth of data from photoproduction at low virtuality
$Q^2$ of the exchanged photon, in particular on jet production
\cite{Klasen:1994bj}, giving access
also to the distributions of partons in the photon \cite{Albino:2002ck}
and to measurements
of the strong coupling constant \cite{Klasen:2014}.

Using the data set of the HERA-II phase of the HERA collider
from 2003-2007 with an integrated
luminosity of $\unit{351}{pb^{-1}}$, the H1 collaboration have recently published
final measurements of inclusive jet, dijet and three-jet production in deep-inelastic
scattering (DIS)
\cite{Andreev:2014wwa,Britzger:2013}
and used them to determine the strong coupling constant (at the mass $M_Z$ of
the $Z$-boson) to be
\beq
 \alpha_s(M_Z)=0.1185\pm0.0016 ({\rm exp.})\pm0.0040\, ({\rm th.}),
 \label{eq:1}
\eeq
taking into account absolute double-differential inclusive jet, dijet and three-jet cross section
data as functions of $Q^2$ and the jet transverse momentum $p_T$. A more precise value was
obtained from normalized jet cross sections, yielding
\beq
 \alpha_s(M_Z)=0.1165\pm0.0008  ({\rm exp.})\pm0.0038\, ({\rm th.})\, .
 \label{eq:2}
\eeq
Unsatisfactorily, only the value obtained by unrenormalized results is in agreement with the
current world average
of $\alpha_s(M_Z)=0.1185\pm0.0006$ \cite{Agashe:2014kda}.
The latter uses only observables that
are known to next-to-next-to-leading order (NNLO) of perturbative QCD, while
the analysis of the H1 collaboration was done in next-to-leading order (NLO) accuracy.
The lack of knowledge of higher-order contributions becomes manifest in a bigger
theoretical uncertainty due to scale variation in Eq.\ (\ref{eq:1}).
The absolute double-differential cross section measurement of inclusive jets
alone led H1 to a value of the strong coupling constant of
\beq
 \alpha_s(M_Z)=0.1174\pm0.0022 ({\rm exp.})\pm0.0050\, ({\rm th.}).
 \label{eq:3}
\eeq

In this Letter, we compute the inclusive jet production DIS cross
section for the first time including NNLO contributions, obtained
from a unified threshold resummation formalism \cite{Kidonakis:2003tx},
and extract an approximate NNLO (aNNLO) value for the strong coupling constant.
Our calculations are based
on previous work on inclusive jet production in deep-inelastic scattering up to NLO
\cite{Potter:1997}. They reduce, as we will see,
the theoretical prediction and increase the central fit value of the strong
coupling constant, improving
the description of the final HERA data in particular at high photon virtuality $Q^2$.

\section{NNLO contributions to jet production in DIS}
\label{sec:2}

The QCD factorization theorem allows to write the differential cross section for inclusive
jet production in neutral-current DIS with high momentum transfer $Q^2=-q^2$
as a convolution of the partonic cross section $\d\sigma_{\gamma a}$
with the parton densities in the proton $f_{a/P}$ and the flux of photons in electrons
$f_{\gamma/e}$ as
\begin{align}
\d\sigma=&\sum_a\int\d y\,f_{\gamma/e}(y) \nonumber \\
&\times\int\d x_P \,f_{a/P}(x_P,\mu_F)\d\sigma_{\gamma a}(\alpha_s,\mu_R,\mu_F) \, ,
\end{align}
where we define $y=(p\cdot q)/(p\cdot k)$ with $p$ and $k$ the momenta of the
incoming proton and electron, respectively, and $q$ the momentum of the exchanged photon.
In deep-inelastic scattering the highly off-shell photon has no time to decay, 
so resolved photon contributions can safely be neglected.

From a unified threshold resummation formalism a master formula
can be obtained that permits to compute soft and virtual corrections
to arbitrary partonic hard scattering cross sections \cite{Kidonakis:2003tx}.
At NLO it reads
\beq
 \d\sigma_{ab}=\d\sigma_{ab}^B{\alpha_s(\mu)\over\pi}
 \le c_3D_1(z)+c_2D_0(z)+c_1\delta(1-z)\re \, ,
\eeq
where for just one color-charged parton in the initial state we only need the formula
for simple color flow. The functions
\bea
 D_l(z)&=&\le{\ln^l(1-z)\over 1-z}\re_+
\eea
with decreasing $l$ are the leading and subleading logarithms at
partonic threshold ($z\to1$) in pair-invariant-mass kinematics.
The NNLO master formula is given in the reference cited above,
as are the general formul\ae\ for the
coefficients $c_i$.

We state here the coefficients for the two partonic processes that contribute to jet
production in DIS. For $\gamma^*q\rightarrow qg$, where $\gamma^*$ represents the off-shell
photon, $g$ a gluon and $q$ a quark or an anti-quark, we find
\beq
 c_3=C_F-N_C \, ,
\eeq
\begin{align}
 c_2 =& \,2C_F \ln{\left(\frac{-u}{M^2}\right)}+N_C\ln{\left(\frac{t}{u}\right)} \nonumber \\
      &\qquad\quad-C_F\ln{\left(\frac{\mu_F^2}{M^2}\right)}-\frac{3}{4}C_F-\frac{\beta_0}{4}
\end{align}
and $c_1=c_1^\mu+T_1$ with
\beq
 c_1^\mu=-\frac{3}{4}C_F\ln{\left(\frac{\mu_F^2}{M^2}\right)}
 +\frac{\beta_0}{4}\ln{\left(\frac{\mu_R^2}{M^2}\right)} \, .
\eeq
For the second process $\gamma^* g \rightarrow q \bar{q}$ we find
\beq
 c_3=2\left( N_C-C_F\right) \, ,
\eeq
\beq
 c_2 = N_C \ln{\left(\frac{tu}{M^4}\right)}-N_C\ln{\left(\frac{\mu_F^2}{M^2}\right)}
 -\frac{3}{2}C_F \, ,
\eeq
and 
\beq
 c_1^\mu = \frac{\beta_0}{4}\left[ \ln{\left(\frac{\mu_R^2}{M^2}\right)}
 -\ln{\left(\frac{\mu_F^2}{M^2}\right)} \right] \, .
\eeq
These coefficients agree with those found in photoproduction for massless jets for the direct
part
\cite{Klasen:2014}. For massive jets, additional logarithms depending on the jet
radius $R$ appear (e.g.\ in $c_2$) \cite{deFlorian:2007fv}, which are however
irrelevant in the case $R=1$ as in the H1 analysis considered here \cite{Andreev:2014wwa}.
The calculation is analogous to the case of single-jet production in hadron-hadron
collisions \cite{Kidonakis:2000gi}.
The above coefficients further depend on the
QCD color factors $C_F=4/3$ and $N_C=3$, on the one-loop
$\beta$-function $\beta_0=(11N_C-2n_f)/3$ with $n_f$ the number of active quark flavors,
the Mandelstam variables $t$ and $u$, the renormalization and factorization
scales $\mu_R$ and $\mu_F$ and the fixed large invariant scale $M^2$, that in DIS
is equal to $Q^2$, whereas in photoproduction it was the Mandelstam variable $s$. The part
$T_1$ of $c_1$ does not contain any dependence either on the renormalization or the
factorization scale. It includes the NLO virtual corrections and is not predicted by
the threshold resummation formalism. If available, it can be read off from a full NLO
calculation. For the case of transverse photon polarization it could be found in reference
\cite{Potter:1997}. For longitudinal polarization we took the formula directly from the
source code of the corresponding program JetViP \cite{Klasen:1998,Potter:2000}, that
calculates inclusive jet production in DIS to NLO accuracy. Some two-loop quantities
appearing in the NNLO master formula were not given explicitly in \cite{Kidonakis:2003tx}
and could not been found in the respective literature. As in our previous work on jet
photoproduction \cite{Klasen:2014}, they were therefore neglected.

\section{Comparisons to H1 data}
\label{sec:3}

The NNLO contributions have been implemented in the code JetViP for inclusive jet and dijet
production in DIS, where the convolution over $z$ was already included for NLO initial-state
corrections on the proton side. At NLO, we use of course our complete calculation and not
only the logarithmically enhanced terms described above. As a numerical check, we have
repeated the NLO analysis of inclusive single-differential jet production of the H1 collaboration,
performed with NLOJet++ \cite{Nagy:2001xb} and
presented in reference \cite{Andreev:2014wwa,Britzger:2013}, and found excellent agreement,
confirming previous successful comparisons of different NLO programs for jet production
in DIS \cite{Potter:1999gg,Nagy:2001xb}.

The measurement took place during the HERA-II running period with an integrated luminosity of
$\unit{351}{pb^{-1}}$. The beam energies were $\unit{27.6}{GeV}$ for electrons or positrons and
$\unit{920}{GeV}$ for protons, which gives a center-of-mass energy of $\unit{319}{GeV}$. The
leptonic phase space was given by $\unit{150}{GeV^2}<Q^2<\unit{15\,000}{GeV^2}$ and $0.2<y<0.7$.
The jet phase space was restricted to the rapidity interval $-1.0<\eta_{lab}<2.5$, where
$\eta_{lab}$ is the pseudorapidity of a jet in the HERA lab frame. The cross section was
measured differentially in the jet transverse momentum $p_T$ and the virtuality $Q^2$.
Jets were reconstructed using the $k_T$-clustering algorithm \cite{Catani:1993hr} in the
Breit frame, where exlusively electroweak processes can be ruled out by demanding a minimum
of jet transverse momentum (here $p_T>\unit{7}{GeV}$). In inclusive jet production in DIS,
an almost identical fit result for $\alpha_s(M_Z)$ was obtained with the anti-$k_T$ algorithm.
The jet radius was $R=1$. The perturbative scales were chosen to be
\beq\label{eq:scales}
 \mu_R^2=(Q^2+p_T^2)/2 \quad \text{and} \quad \mu_F^2=Q^2\, .
\eeq
The perturbative hard-scattering functions were convoluted with the MSTW2008 set of parton
distribution functions in the proton with different fixed $\alpha_s(M_Z)$ values
\cite{Martin:2009}. This PDF set especially offers the possibility to determine a best-fit
$\alpha_s(M_Z)$. The number of active flavors was $n_f=5$, since sea contributions of a heavy
top quark inside the proton can safely be neglected. Following the H1 analysis, we choose the
PDF member with $\alpha_s(M_Z)=0.118$ in all plots shown in this Letter, which together with
the scale choice from Eq.\ (\ref{eq:scales}) defines our central fit.

%
\begin{figure}
\centering
\includegraphics[width=\columnwidth]{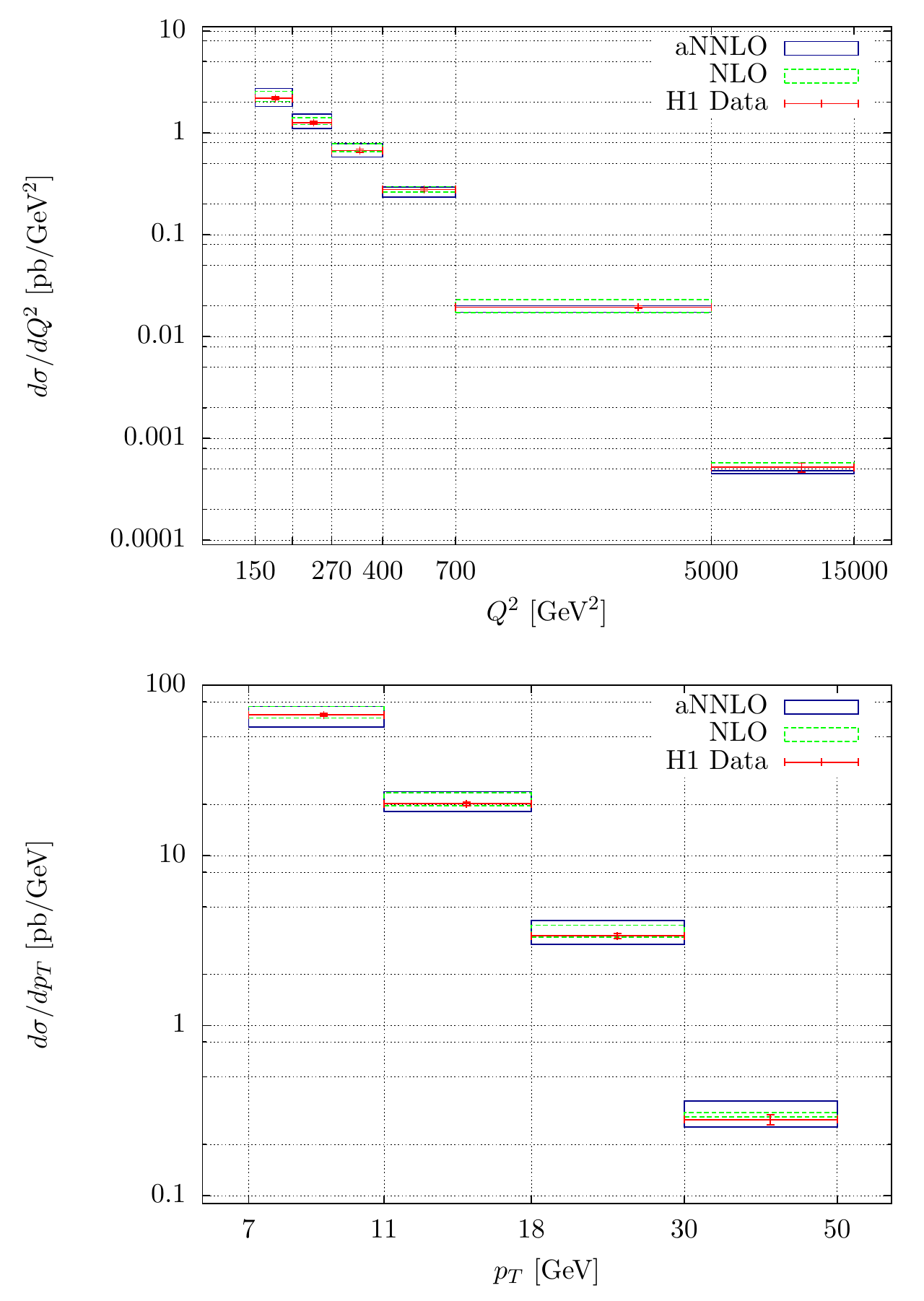}
\caption{\label{fig:nnloscv}Single-differential inclusive jet cross sections
  as a function of photon virtuality $Q^2$ (top) and jet transverse momentum $p_T$
 (bottom) in NLO (green dashed lines) and aNNLO (blue full lines) with the corresponding
 scale uncertainty bands, obtained by varying $\mu_R$ and $\mu_F$ simultaneously by a
 factor of two up and down, compared to the final H1 data (red points, color online).}
\end{figure}
%

In Fig.\ \ref{fig:nnloscv}, we compare our NLO (green dashed lines) and aNNLO (blue full lines)
results to the experimental data of the H1 collaboration (red points). The uncertainty bands are
obtained by varying both scales simultaneously
about the central scales up and down by a factor of two as it was also done by the H1
collaboration. We have verified that our NLO results and uncertainty bands agree very well with
those of the H1 analysis. Comparing our calculations to the measurements, we see that the data
points lie in the error bands of both the NLO and the aNNLO calculations.
For the $Q^2$ distribution (top), the scale uncertainty is of similar size at NLO and aNNLO
at low $Q^2$ and is considerably reduced in aNNLO at higher $Q^2$ as expected.
We checked that this tendency continues and becomes more pronounced for even higher $Q^2$,
where unluckily we have no data to compare with.
For the $p_T$ distribution (bottom), the scale uncertainty is not reduced from NLO to aNNLO,
since analytically no logarithms of ratios of $\mu_R$ or $\mu_F$ over $p_T$ appear
(see above). While in our previous analysis of jet photoproduction the jet transverse momentum
$p_T$ ranged from 17 to 71 GeV \cite{Klasen:2014}, its range is restricted in DIS and this
analysis to lower values of 7 to 50 GeV, while the photon virtuality reaches values up to
$(122.5$ GeV)$^2$.

%
\begin{figure}
\centering
\includegraphics[width=\columnwidth]{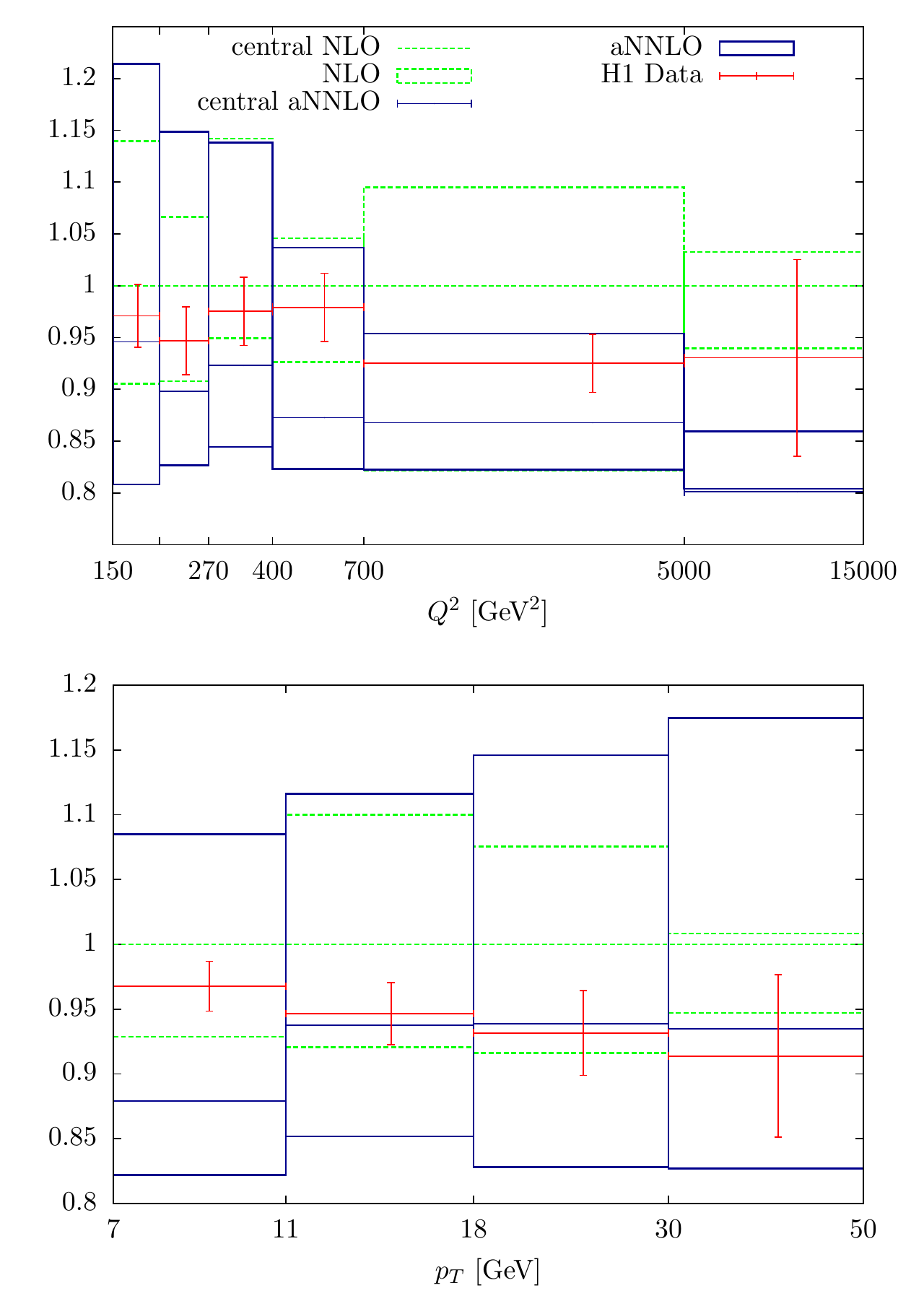}
\caption{\label{fig:nnloscvn}Same as Fig.\ \ref{fig:nnloscv}, but normalized
 to the central NLO predictions.}
\end{figure}
%

In Fig.\ \ref{fig:nnloscvn}, we show the same comparison normalized to the central NLO result.
We also depict here the central aNNLO result, which is always smaller than the NLO central result
by approximately 6\%.
The central $Q^2$ distribution and even more the central $p_T$ distribution agree better with
the H1 data at aNNLO than at NLO, as they then lie right within the experimental uncertainties.
While the central NLO results overestimate the measured cross sections, indicating that the
value of the strong coupling constant $\alpha_s(M_Z)$ used at this order is too large, the
central aNNLO results underestimate the data and require a slightly larger value of $\alpha_s$.
The data are thus clearly sensitive to the strong coupling constant and can be used for an
extraction not only at NLO, but also at aNNLO. However, we do not expect a significant
reduction of the theoretical error from the scale uncertainty, in particular from the $p_T$
distribution.

\section{Determination of {$\boldmath{\alpha_s}$}}
\label{sec:4}

To determine the strong coupling constant from these comparisons, the
theoretical calculations have to be performed with a set of parton
densities in the proton obtained from global fits assuming different
values of $\alpha_s(M_Z)$. For our analysis at aNNLO, we employ the
latest fits of the MSTW group, which have been obtained
with NNLO running of the coupling, evolution of the parton densities, deep-inelastic
scattering and vector-boson production matrix elements \cite{Martin:2009}. 
22 different MSTW2008 NNLO members were used, which correspond to values of
$\alpha_s(M_Z)=0.107$ to 0.127. To compare the aNNLO best-fit $\alpha_s$ to a
corresponding NLO $\alpha_s$ based on our full NLO calculation, we carry out the same
approach with the NLO MSTW2008 parton distribution functions. These are available for the
range of $\alpha_s(M_Z)=0.110$ to 0.130. 

The strong coupling constant $\alpha_s$ was determined by comparing the theoretical predictions
at NLO and aNNLO to the experimental measurements by H1 and then finding the minimum value of the
reduced $\chi^2$. We present here our results for the single-differential $p_T$-distribution, but
have verified that fitting the $Q^2$-distribution yields very similar results. At NLO we find
\beq
 \alpha_s^{\rm NLO}(M_Z)=0.115\pm0.002({\rm exp.})\pm0.005({\rm th.}),
\eeq
where the central value is slightly lower than the one obtained by H1 from the unnormalized
double-differential inclusive jet cross section (cf.\ Eq.\ (\ref{eq:3})), but where the
total experimental error and the theoretical error obtained from a simultaneous variation of
the renormalization and factorization scales agree very well. At aNNLO, $\alpha_s$ gets shifted
upwards, since we demonstrated above that the aNNLO contributions reduce the differential cross
sections compared to NLO. We find
\beq
 \alpha_s^{\rm aNNLO}(M_Z)=0.122\pm0.002({\rm exp.})\pm0.013({\rm th.}),
\eeq
where the central value is now slightly above the world average, the experimental error
is of course unchanged, and the theoretical error is slightly larger, reflecting the
observation made above that the aNNLO calculation is not yet sufficiently stabilized by threshold
logarithms at these values of $p_T$ and $Q^2$. The numerical situation would only
improve at higher values of $Q^2$, where unfortunately no experimental data are
available.

\section{Conclusions}
\label{sec:5}

In conclusion, we have presented here a first calculation of inclusive-jet production in
neutral-current deep-inelastic scattering up to NNLO of perturbative QCD.
Leading and subleading logarithmic contributions were extracted from
a unified threshold resummation formalism for virtual photon-parton scattering processes
and shown to agree with those appearing in our full NLO calculations.
The aNNLO contributions implemented in our NLO program improve the description of the
final H1 data on inclusive-jet production in the $Q^2$ distribution and even more in the
$p_T$ distribution, when the world average value of $\alpha_s$ is used and for central
scale choices. The scale uncertainties are reduced only at the highest values of $Q^2$,
where threshold corrections are most important. An aNNLO fit of these data with the
MSTW2008 set of parton densities resulted in a new determination of the strong coupling
constant at the mass of the $Z$-boson that increased the central fit value from below
to above the current world average, but did not reduce the theoretical error.


We thank W.\ Vogelsang for useful discussions.
This work has been supported by the BMBF Theorie-Verbund ``Begleitende
theoretische Untersuchungen zu den Experimenten an den Gro\ss{}ger\"aten der
Teilchenphysik.''


\end{document}